\newcommand{\kms}{{~\rm km\; s^{-1}}}
\newcommand{\cc}{{~\rm cm^{-3}}}

\newcommand{\s}{{~\rm s}}
\newcommand{\km}{{~\rm km}}
\newcommand{\g}{{~\rm g}}
\newcommand{\gcm}{{~\rm g }{~\rm cm}^{-3}}
\newcommand{\K}{{~\rm K}}
\newcommand{\erg}{{~\rm erg}}
\newcommand{\yr}{{~\rm yr}}
\newcommand{\Myr}{{~\rm Myr}}

\newcommand{\kpc}{{~\rm kpc}}

\documentclass[12pt,preprint,a4paper]{aastex}
\usepackage{amsmath}
\shortauthors{Sternberg et al.}

\begin{document}

\title{EXPLAINING THE ENERGETIC AGN OUTBURST OF MS0735+7421 WITH MASSIVE
SLOW JETS}

\author{Assaf Sternberg\altaffilmark{1},  Noam Soker\altaffilmark{1}}

\altaffiltext{1}{Department of Physics,
Technion$-$Israel Institute of Technology, Haifa 32000, Israel;
phassaf@techunix.technion.ac.il; soker@physics.technion.ac.il}

\begin{abstract}
By conducting axisymmetrical hydrodynamical numerical simulations (2.5
dimensional code) we show that slow, massive, wide jets can reproduce the
morphology of the huge X-ray deficient bubble pair in the cluster of galaxies
MS0735+7421. The total energy of the jets, composed of the energy in the
bubble pair and in the shock wave, is constraint by observations conducted
by McNamara et al. (2009) to be $\sim 10^{62} \erg$. We show that two opposite
jets that are active for $\sim 100~$Myr, each with a launching half opening
angle of $\alpha \simeq 70^\circ$, an initial velocity of $v_j \sim 0.1 c$,
and a total mass loss rate of the two jets of
$\dot M_{2j} \sim 100 M_\odot \yr^{-1}$, can account for the observed
morphology. Rapidly precessing narrow jets can be used instead of wide jets.
In our model the cluster suffered from a cooling catastrophe $\sim 100~$Myr
ago. Most of the mass that cooled, $\sim 10^{10} M_\odot$, was expelled back
to the intracluster medium (ICM) by the AGN activity and is inside the bubbles
now, $\sim 10\%$ formed stars, and $\sim 10\%$ of the cold gas was accreted
by the central black hole and was the source of the outburst energy. This type
of activity is similar to that expected to occur in galaxy formation.
\end{abstract}

{{\it Subject headings:}
galaxy clusters: cooling flows $-$Active Galactic Nuclei: individual
(MS0735.6+7421).

\section{INTRODUCTION}
\label{sec:intro}

Many clusters of galaxies harbor bubbles (cavities) devoid of X-ray emission,
e.g., Hydra A (McNamara et al. 2000), A 2597 (McNamara et al. 2001), RBS797
(Schindler et al. 2001), Abell 496 (Dupke \& White 2001), Abell 4059 (Heinz
et al. 2002), Perseus (Fabian et al. 2000) and Abell 2052 (Blanton et al.
2001, 2009). These low density bubbles are inflated by jets launched by the active
galactic nuclei (AGN) sitting at the centers of these clusters. The properties
of these jets are poorly understood. Although in most cases radio jets are
observed to be associated with X-ray deficient bubbles, there is no general
one-to-one correspondence between the radio jets and bubbles. In Abell 2626
(Rizza et al. 2000; Wong et al. 2008) and in Hercules A (Nulsen et al. 2005;
Gizani \& Leahy 2003), for example, no X-ray deficient bubbles are observed
at the locations of strong radio emission. \par

The bubbles we are aiming to understand have their width not much smaller,
and even larger, than their length; we term them `fat bubbles'. In previous
papers (Sternberg et al. 2007; Sternberg \& Soker 2008a,b) it has been shown
that large bubbles can be inflated by slow-massive-wide (SMW) jets. Typical
values are $v_j \sim 10^4 \km \s^{-1} \sim 0.02-0.1c$, for the
initial jet speed, $\dot M_{2j} \sim 1-50 M_\odot \yr^{-1}$ for the mass loss
rate into the two jets, and a half opening angle of $\alpha \ga 30^\circ$.
Precessing narrow jets have the same effect as wide jets (Sternberg \& Soker
2008a). In contrast to these simulated wide jets,
synchrotron emission, mainly in radio and in synchrotron self-Compton X-ray and
optical emission (e.g., Cygnus A, Wilson et al. 2000), show 
relativistic jets
to be narrow. For that, it was argued, e.g., Binney (2004), that in many
cases the relativistic jets carry a small fraction of the mass and energy in
the outflow. \par

In this paper we continue with the view held in our previous papers, that
radio jets do not necessarily trace the main energy and mass carried by the
outflow from the AGN vicinity, neither in the period of activity nor in the outflow
geometry. We apply this conjecture to the energetic outburst of the cluster
MS0735.6+7421 (hereafter MS0735), where the bright radio jets reside inside
the bubbles, but do not follow their exact geometry (McNamara et al. 2005).
The very large bubbles in this cluster demand that the jets which inflated 
them were
extremely energetic (Gitti et al. 2007). The lack of strong AGN emission and
a low upper limit on present star formation rate, motivated McNamara et al.
(2009) to propose that the jets were powered by the rotational energy (spin)
of the central black hole (BH). They cite the model of Meier (1999, 2001),
in which the dominant energy source of the jets is claimed to be the spinning
BH. However, it is questionable whether the energy supplied by the spinning
BH can exceed by much (or at all) the energy supplied by the accretion process
(Livio et al. 1999; Li 2000) \par

{{{
We use only wide (or precessing jets) because narrow jets expand to large
distances before they inflate large bubbles. This is clearly seen in figure 1 of
Sternberg et al. (2007), where a jet with a half opening angle of $20^\circ$ inflates
a long narrow bubble rather than a more spherical bubble.
Analytical arguments for wide jets are given in Soker (2004).
In figure 2 of Sternberg \& Soker (2008a) it can be seen that when a narrow jet precesses
too close to the symmetry axis ($15^\circ$ in that case), it inflates a long bubble, rather
than a large one as desired here.
}}}

In this paper we propose that the bubbles in MS0735 were inflated by SMW, or
precessing, jets in an outburst that has ceased $t_{\rm Q}$ years ago, where
$t_{\rm Q}$ is of the order of a few $10^7 \yr$. The energy came
from the
accretion process and not from the spinning BH. The present radio activity
is a weak leftover from that eruption. The proposed scenario is outlined in
$\S 2$, and implication for feedback in galaxy formation is discussed. The
numerical method and initial conditions are described in $\S 3$, and the
results of the numerical simulations are presented in $\S 4$. Our discussion
and summary are in $\S 5$.

\section{THE ENERGETIC OUTBURST OF MS0735}
\label{sec:outburst}

The bubble pair in MS0735 that we try to model is presented in Figure 
\ref{fig:image} (taken from McNamara et al. 2009, with some scaling added).
\begin{figure}
\hskip 2.0 cm
{\includegraphics[scale=0.8]{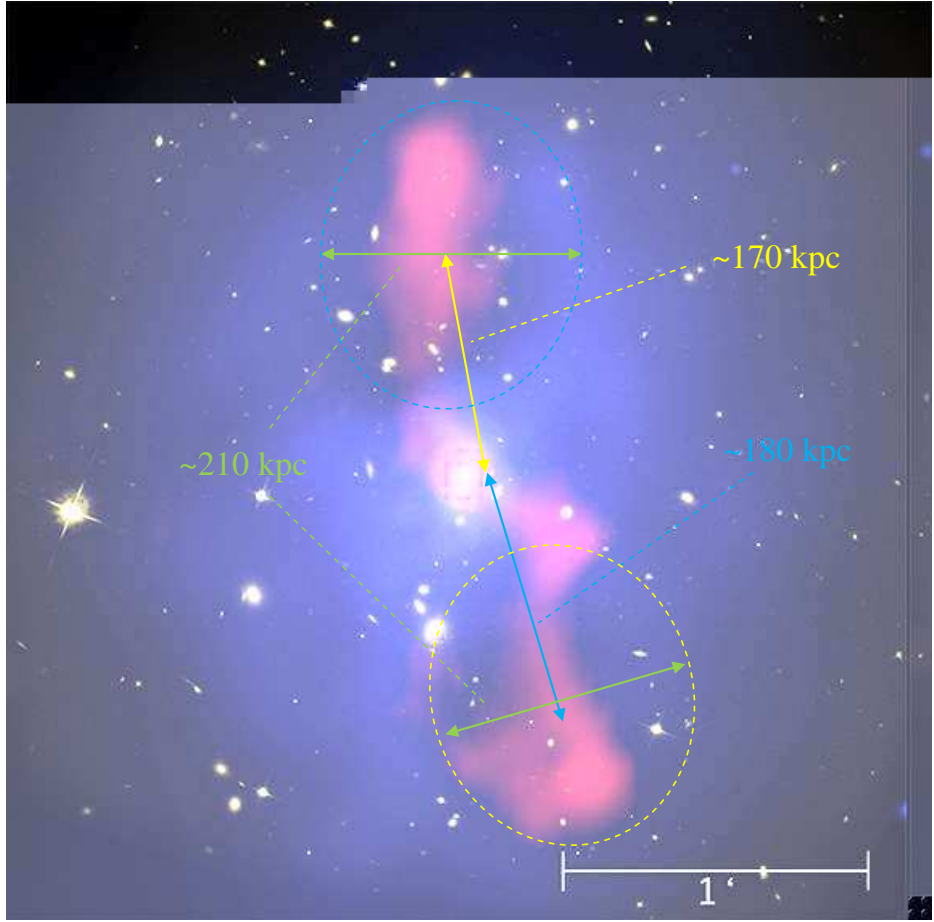}}
\caption{X-ray image of the bubble-pair in MS0735.6+7421 cluster (from McNamara et al. 2009).
Image of the inner 200 arcsec (700 kpc) combining
the X-ray (blue), I-band (white), and radio wavelengths (red).
The two-sided arrow lines with different lengths and the dashed ellipses are 
added to allow comparison with the numerical results.}
\label{fig:image}
\end{figure}

McNamara et al. (2009) estimated the total energy of the outburst in MS0735,
i.e, both in the bubbles and the shocks, to be $E_{2j}=1.2 \times 10^{62} \erg$,
{{{about half of it in the bubble and the rest in the outer shock wave. }}}
This estimate is based on the jet material being relativistic,
i.e., $\gamma=4/3$. For jets of non-relativistic material, i.e., $\gamma=5/3$,
the {{{ energy in the bubble is $\sim 4 \times10^{61}~\erg$, }}}
and the total energy estimate is $E_{2j} \simeq 10 \times10^{61}~\erg$.
{{{ We will run jets with a total energy of $ E_{2j} = 7.2 \times10^{61}~\erg$. }}}
For an efficiency of $\epsilon=0.1$ of converting accretion energy to jets'
power, the total accreted mass required to inflate the bubbles is
$M_{\rm acc} \simeq 4 \times 10^8 M_\odot$.
{{{ This amount of accreted mass holds even if the jets are relativistic, as long as one does
not make use of the energy stored in the spin of the BH. }}}
To account for this energy and accreted mass, we propose that the high activity
(energetic) phase lasted for a
time period of $\Delta t_{\rm H} \sim 10^8 \yr$ that has ended
$t_{\rm Q}$ (a few $10^7$) years ago. During the time period
$t_{\rm Q}$ the AGN activity has continued, but with a much lower power, about
equal to the present power of the radio jets, and no star formation took
place. In our model the jets that inflated the bubbles$-$the inflating
jets$-$were slow, i.e., highly non-relativistic with $v_j \simeq 0.1c$, and
massive, i.e., the mass outflow rate was much higher than the mass accretion
rate. Feedback heating of cooling flow clusters with massive slow jets was
studied before (Pizzolato \& Soker 2005a; Soker \& Pizzolato 2005; Sternberg
et al. 2007). The total mass carried by the two jets is derived from the assumption that
the kinetic energy of the jets is converted to the total energy in the bubbles and shocks,
and it gives
\begin{equation}
M_{2j} = 8.4 \times 10^{9}
\left( \frac {E_{2j}}{7.5 \times 10^{61} \erg} \right)
\left( \frac {v_j}{0.1 c} \right)^{-2} M_\odot .
\label{eq:massjet}
\end{equation}
We emphasize that in \emph{mass loss rate in the jets} or \emph{total mass
carried by the jets} we mean both the jet material and the material advected
onto it.
Basically, the BH itself ejected relativistic jets. However, within a short
distance from the center, $r < 1 \kpc$, they interacted with the dense
cooling gas, and formed SMW jets (Soker 2008). \par

Based on a UV power of $1.82 \pm 0.14 \times 10^{42} \erg \s^{-1}$ in the
$1800-2600$\AA\ band,  McNamara et al. (2009) deduced a present day star
formation rate of $\dot M_{\rm present} = 0.25 M_\odot \yr^{-1}$. To estimate
the mass that was converted to stars $\Delta t_H+t_Q \simeq 10^8 \yr$ ago we
use figure 1 from Fioc \& Rocca-Volmerange (1997).
{{{ In that figure Fioc \& Rocca-Volmerange (1997) give the spectral evolution
after an instantaneous burst of star formation.    }}}
The $1800-2600$\AA~specific flux at $10^8 \yr$ after a star formation burst
can be fitted with
$F_\lambda = 7 \times 10^{30} (\lambda / 2000$\AA$)^{-3/2} \erg \s^{-1}
$\AA$^{-1} M_\odot ^{-1}$.
Integrating in the $1800-2600$\AA ~band, we find that a flux of
$1.82 \pm 0.14 \times 10^{42} \erg \s^{-1}$ implies a total star formation
mass of $3.7 \times 10^8 M_\odot$ (assuming a negligible present star
formation rate) $10^8 \yr$ ago. As the star formation rate was continuous and 
ended
$\sim 10^8 \yr$ ago, we can take the average star formation time to be
$\ga 1.5 \times 10^8 \yr$ ago, and the total stellar mass that formed
in the outburst to be larger $M_{\rm stars} \sim 10^9 M_\odot$. \par

We use the above values to propose the following scenario for the inflation
of the bubbles in MS0735. For a time period of $\sim 10^8 \yr$, a cooling rate of
$\sim 100 M_\odot \yr^{-1}$ result in a mass of $\sim 10^{10} M_\odot$ flowing toward
the center of MS0735.
Some of the cool mass might have come from a merger event. However, we basically
propose that MS0735 went through a phase that might be termed a cooling
catastrophe. The strong AGN activity ejected most of the cooling mass before
much star formation occurred. A vigorous termination of star formation is
needed in galaxy formation as well, where the expelled mass can be $\sim 10$
times the mass that converted to stars (e.g., Bower et al. 2008).
The process we propose here might resemble the process that terminated star
formation after the main phase of galaxies assembly. \par

There are three possible reasons that make the accretion and jet ejection that
took place during the outburst of M0735 different than those during the main
assembly phase of galaxies and their central BH. First, in MS0735 the central
BH is an ultra-massive one (McNamara et al. 2009). This might lead to a high
accretion rate that results in very strong jets, that ejected a large fraction
of the cooling gas back to large radii. Second, the cooling catastrophe
occurred after a hot dense ICM was built up in the cluster, such that large
quantities of gas could cool, disturb the relativistic jets, and form
slower, denser, and wider jets (Soker 2008). Third, most likely there is a
massive BH companion to the central BH of MS0735 (Pizzolato \& Soker 2005b).
The evidence for precession is the point symmetric structure of the lobes.
Pizzolato \& Soker (2005b) considered a case of a BH companion of a mass
$\sim 10 \%$ the mass of the central BH, and at a distance of a few~pc. \par

A small fraction, $\sim 5 \%$, of the cooling mass was accreted by the BH,
summing to a total of $M_{\rm acc} \simeq 4 \times 10^8 (\epsilon/0.1) ^{-1} M_\odot$,
and a similar mass, $M_{\rm stars} \sim 10^9 M_\odot$, was converted to stars.
However, the huge amount of cool material near the center disturbed the relativistic jets
blown by the BH within a short distance from the center, $< 1 \kpc$. The mass
that disrupted the propagation of the jet was $\ga 100$ times larger than the
mass in the primary relativistic jets.
This condition leads to an efficient transfer of energy from
the relativistic jets to the disrupting gas, and the formation of slow massive
wide (SMW) jets (Soker 2008). The material in the massive jets resides now
inside the two bubbles, after it was heated when it was shocked. Both mass
and energy are ingredients in the feedback heating of this
cluster, as was suggested in the moderate cooling flow model (Pizzolato \&
Soker 2005a; Soker \& Pizzolato 2005; Soker 2006). \par

The parameters of the proposed model are summarized in Table 1. Some of these
parameters are used explicitly in the numerical simulations to be described in the
next section (they are marked by an asterisk).
With several numerical trials not presented here, we found that an active phase duration of
$\sim 10^{8} \yr$ can reproduce the observations. We took a round number of $10^8 \yr$.
Much longer active phase will result in bubbles that have time to buoy to too large
distances, and a too short active phase requires very high mass outflow rate,
that pushes its way too fast along the symmetry axis.
We then continue the bubble evolution for a time $t_{\rm Q}$ until it matches observations.
As there are uncertainties in the velocity (hence mass loss rate) of the wide jets
to be used in our model, we simulated three different cases, named Model I , II, and III.
\begin{table}[h]

Table 1: Model Parameters

\footnotesize
\bigskip
\begin{tabular}{|l|c|c|c|c|}
\hline
  &  Symbol & Model I & Model II  & Model III \\
\hline
\hline
Jet's half opening angle$^\ast$  & $\alpha$ & $70^\circ$ & $70^\circ$
& $70^\circ$ \\
Mass loss rate in both jets$^\ast$ ($M_\odot \yr^{-1}$) & $\dot{M}_{2j}$ &
$242.4$ & $60.6$ & $26.9$ \\
Jet's velocity$^\ast$  & $v_j$ &  $0.0576c$ & $0.1152c$ & $0.1727c$   \\
Power of both jets$^\ast$ ($10^{46} \erg \s^{-1}$) & $\dot{E}_{2j}$ & $2.28$ &
$2.28$ & $2.28$  \\
Duration of active phase$^\ast$ &$\Delta t_{\rm H}$ & $10^8 \yr$ & $10^8 \yr$ &
$10^8 \yr$  \\
Time since end of active phase$^\ast$ & $t_{\rm Q}$ & $2.5\times10^7 \yr$ &
$1.5\times10^7 \yr$ & $1.5\times10^7 \yr$ \\
Total injected energy ($10^{61} \erg$) & $E_{2j}$ & $7.2$ & $7.2$ & $7.2$ \\
Total injected mass ($10^9 M_\odot$) & $M_{2j}$ & $24$ & $6.1$ & $2.7$ \\
Total accreted mass by the BH ($10^9 M_\odot$) & $M_{\rm acc}$ & $0.4(\epsilon/0.1)^{-1}$
             &  $0.4(\epsilon/0.1)^{-1}$ &  $0.4(\epsilon/0.1)^{-1}$  \\
Total stellar mass formed ($10^9 M_\odot$) & $M_{\rm stars}$ & $\sim 1$ & $\sim 1$ &
$\sim 1$ \\
\hline
\end{tabular}

\footnotesize
\bigskip
Parameters that are marked by $^\ast$ are those used explicitly in the
the numerical simulations.
$\epsilon$ is the energy production efficiency of the accreting central BH.
\normalsize
\end{table}

Few words are appropriate here about the high mass loss rate we use in the simulations.
In our previous papers we gave the supporting arguments for using such a high
mass loss rate. What we like to emphasize here is that we use wide (or precessing;
Sternberg \& Soker 2008a) jets with high mass loss rates based on physical
considerations (Soker 2004; Sternberg et al. 2007; Sternberg \& Soker 2008a),
and not from numerical limitations. In recent years several papers have used
jets with very high mass loss rates (e.g., Simionescu et al., 2009).
Such high mass outflow rates must come with an appropriate physical 
discussion.

The total BH accreted mass of $\sim 4 \times 10^{8} M_\odot$ is not
a concern, as the BH in MS0735 is super massive,
$M_{\rm BH} \simeq 5 \times 10^{9} M_\odot$,
and possibly even  $M_{\rm BH} \ga 10^{10} M_\odot$ (McNamara et al. 2009).

\section{NUMERICAL METHODS AND INITIAL CONDITIONS}
\label{sec:numerics}

The simulations were performed using the \emph{Virginia Hydrodynamics-I}
code (VH-1; Blondin et al. 1990; Stevens et al. 1992), as described in
Sternberg \& Soker (2008b). In this paper we mention only the important
features of the code. \par
Gravity was included in the form of a constant gravitational potential of
a dark matter halo using the NFW profile (Navarro et al. 1996),
\begin{equation}
\Phi_{NFW}(r)=4\pi r^2_s\delta_c\rho_{crit}G
\left[1-\frac{ln(r/r_s+1)}{r/r_s}\right],
\end{equation}
where $r_s=\frac{r_v}{c}$ is a scale radius, $r_v=1700\kpc$ is the virial 
radius, $c=3.45$ is the concentration factor, 
$\delta_c=\frac{200}{3}\frac{c^3}{ln(1+c)-\frac{c}{1+c}}$, and 
$\rho_{crit}=\frac{3H^2}{8\pi G}$ is the critical density of the Universe 
at $z=0$. The ICM gas was set in hydrostatic equilibrium,
\begin{equation}
\rho_{gas}(r)=\rho_{gas,0}e^{-b}\left(\frac{r}{r_s}+1\right)^\frac{b}{r/r_s},
\end{equation}
with $b\equiv4\pi r^2_s\delta_c\rho_{crit}\mu m_pG/k_BT_v$,  $M_v$ is the 
virial mass, and $T_v=\frac{\gamma G\mu m_pM_v}{3k_Br_v}=5.7\times10^7\K$ is 
the virial temperature, which is the average temperature given by Gitti et al. 
(2007) for the relevant area simulated in our simulations. 
The central density was set to $\rho_{gas,0}=1.9\times10^{-26}\g\cc$. 
Radiative cooling was not included. \par 
We study a three-dimensional axisymmetric flow with a 2D grid (referred to 
as 2.5D). Our
computational domain was $r\in[1,400]\kpc$ and $\theta\in[0,\pi/2]$. Hence,
we simulate a quarter of the meridional plane using the two-dimensional
version of the code in spherical coordinates. The symmetry axis of all plots
shown in this paper is along the $x$ (horizontal) axis. The $y$ axis is the
equatorial plane. Reflective boundary conditions were implemented at the
symmetry axis, i.e., $\theta=0$, and at the equatorial plane, i.e,
$\theta=\pi/2$. At the inner and outer radii an inflow/outflow boundary
condition was implemented, i.e., matter was free to flow in/out of the
computational grid. \par

The point-symmetric morphology of the bubbles in MS0735 suggest that the
inflating jets precessed. However, our 2.5D numerical code can only simulate
precessing jets with a very short rotation period around the symmetry axis.
In our previous papers we have shown that the behavior of these jets is
very similar to the behavior of wide jets.
Therefore, we use wide jets in our simulations, with the understanding that rapidly precessing
jets give similar results.  \par

The jets were injected at a radius of $1 \kpc$, with a constant power of
$\dot{E}_j=1.14\times10^{46}\erg\s^{-1}$ per one jet, and a constant
radial velocity $v_j$, hence with a constant mass flux of $2\dot{E}_j/v_j^2$ per one jet,
within a half opening angle of $\alpha$. See Table 1 for the properties of
the different jets we used.

\section{NUMERICAL RESULTS}
\label{sec:results}

Figure \ref{fig:M15} shows density maps (density given in log scale of $\gcm$)
of the evolution of the model I jet (see Table 1 for the jet's parameters).
The jet was turned off at $t=100\Myr$, at which time a low density cavity with
a radius of $\sim100\kpc$ has been inflated. For clarity, we note
that by \emph{the bubble} we mean the shocked jet material, i.e., the bubble 
does not include the jet material before the reverse shock situated at
$\sim40\kpc$. The bubble continues to rise and buoy outward from the cluster
center. \par

At $t=125\Myr$ the bubble's center resides at a distance of
$\sim170\kpc$ from the cluster center. At this time the bubble is a bit
elongated with a major axis of $\sim 220\kpc$ and a minor axis of
$\sim 180\kpc$. The ratio between the density of the matter inside the bubble,
$\rho_{\rm bub}$, and the density of the surrounding ICM, $\rho_{\rm {ICM}}$, is
$\rho_{\rm bub}/\rho_{\rm {ICM}}\sim0.1-0.01$. The forward shock is elliptical and
located at $\sim330\kpc$ near the symmetry axis and at $\sim240\kpc$ near the
equatorial plane.
{{{ The shock eventually decays into a sound wave. However, at the end of our calculation it
is still a shock wave. }}}
   
The arrows given in the bottom panel of Fig. \ref{fig:M15}
represent the velocities. {{{ The arrows represent four velocity bins: }}}
$(i)$ $0.1c_s\leq v_j\leq c_s$ (shortest arrows);
$(ii)$ $c_s\leq v_j\leq7.5c_s$;
$(iii)$ $7.5c_s\leq v_j\leq15c_s$;
$(iv)$ $15c_s\leq v_j\leq30c_s$ (longest arrows),
where $c_s=1152\kms$ is the unperturbed ICM speed of sound.
Figure \ref{fig:M15_temp} shows the temperature map of this
bubble at the same times. The matter inside the bubble has a temperature of
about an order of magnitude higher then that of the surrounding ICM. The
temperature map taken at $t=100\Myr$ clearly shows the distinction between the
pre-shocked jet material and the shocked jet material, i.e., the bubble. \par


\begin{figure}
\hskip 1.0 cm
\begin{tabular}{c}
{\includegraphics[scale=0.5]{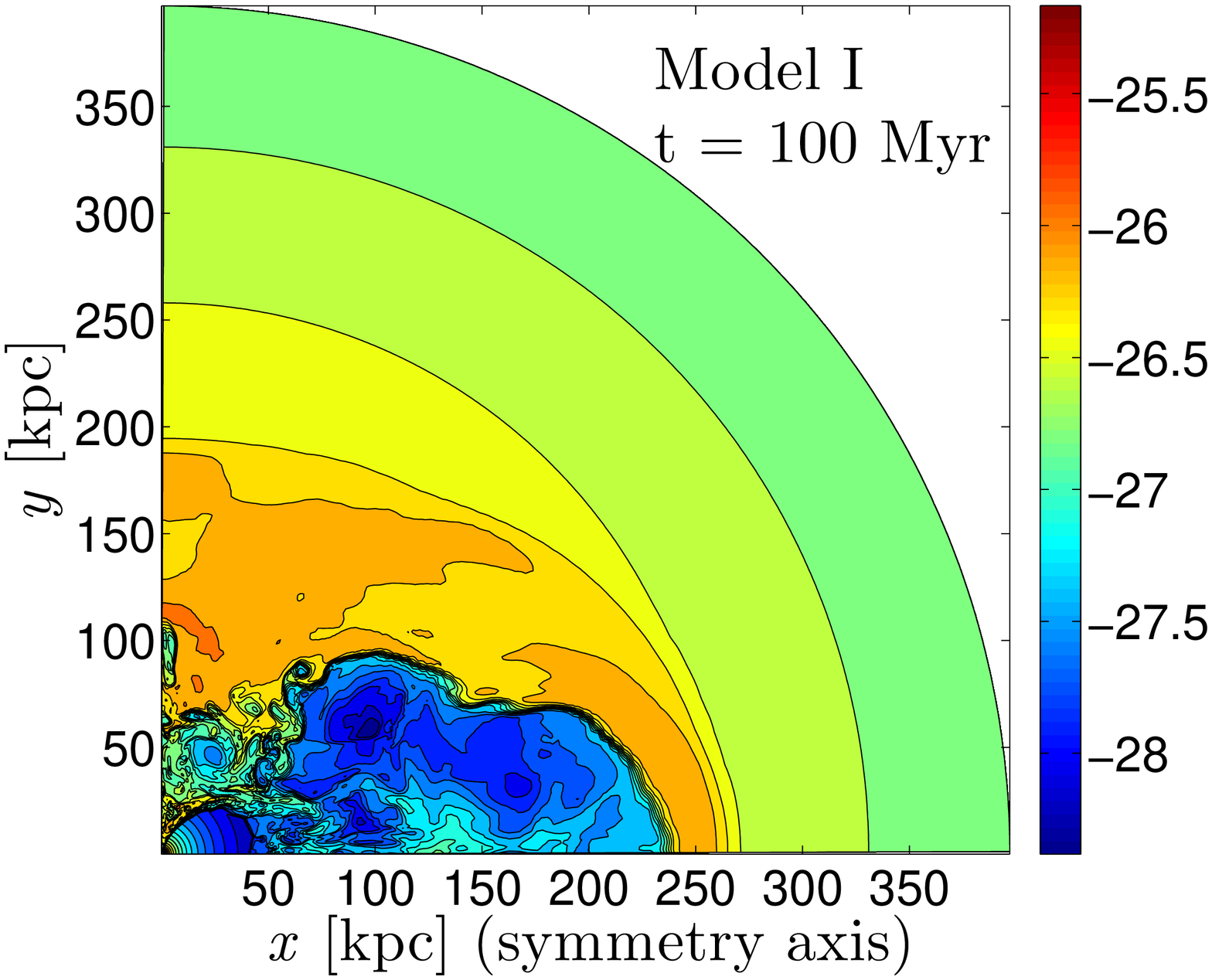}} \\
{\includegraphics[scale=0.5]{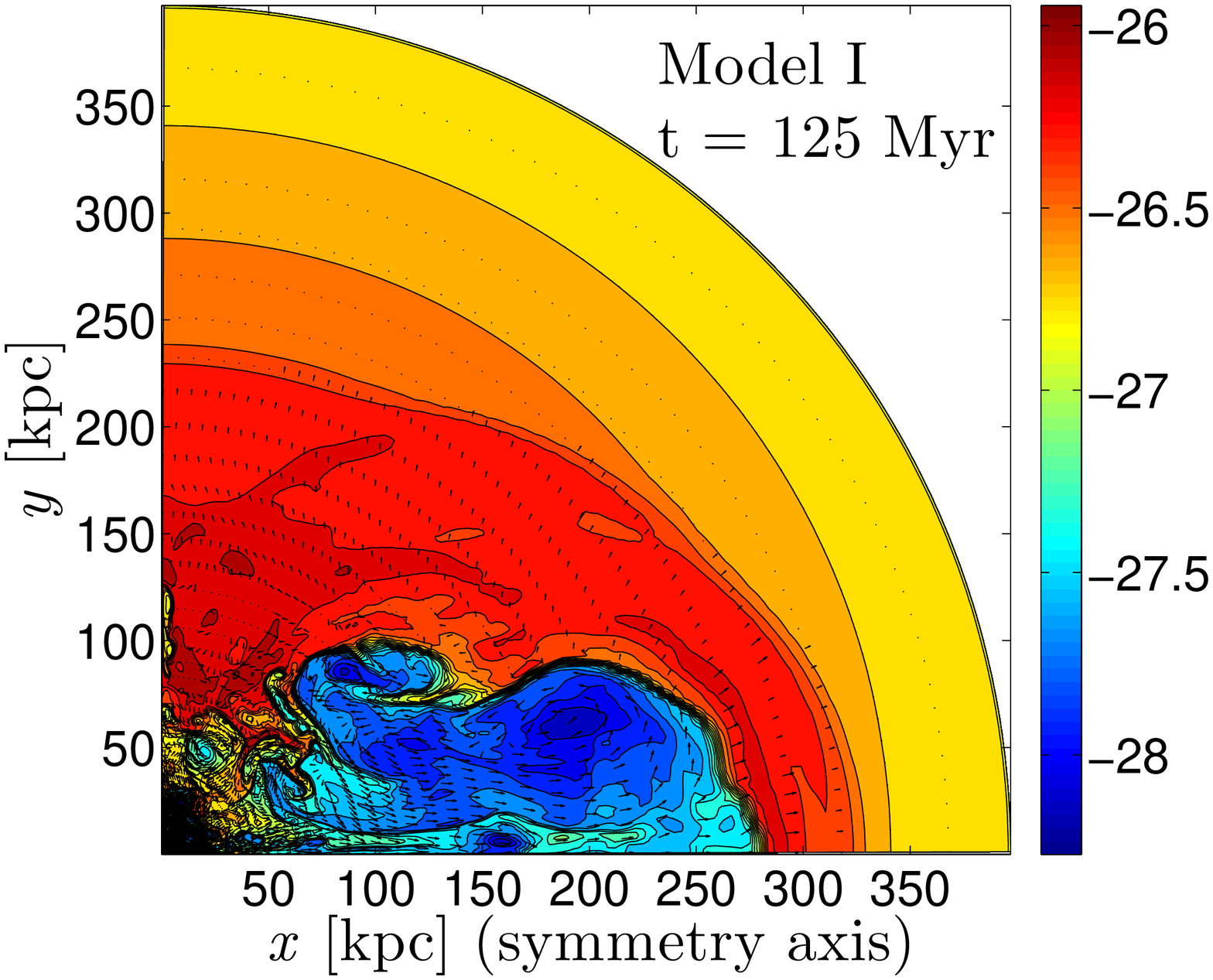}}
\end{tabular}
\caption{Log-density maps of the evolution of the model I jet at $t=100\Myr$
and at $t=125\Myr$, as indicated in the figures. The jet was active for
$100\Myr$ and was turned off at $t=100\Myr$. The density is given in
$\log(\gcm)$.
{{{
The arrows drawn in the bottom panel represent the flow velocity, and
are grouped into four velocity bins:
$(i)$ $0.1c_s\leq v_j\leq c_s$ (shortest arrows);
$(ii)$ $c_s\leq v_j\leq7.5c_s$;
$(iii)$ $7.5c_s\leq v_j\leq15c_s$;
$(iv)$ $15c_s\leq v_j\leq30c_s$ (longest arrows),
where $c_s=1152\kms$ is the unperturbed ICM speed of sound. }}}
}
\label{fig:M15}
\end{figure}

\begin{figure}
\hskip 1.0 cm
\begin{tabular}{c}
{\includegraphics[scale=0.5]{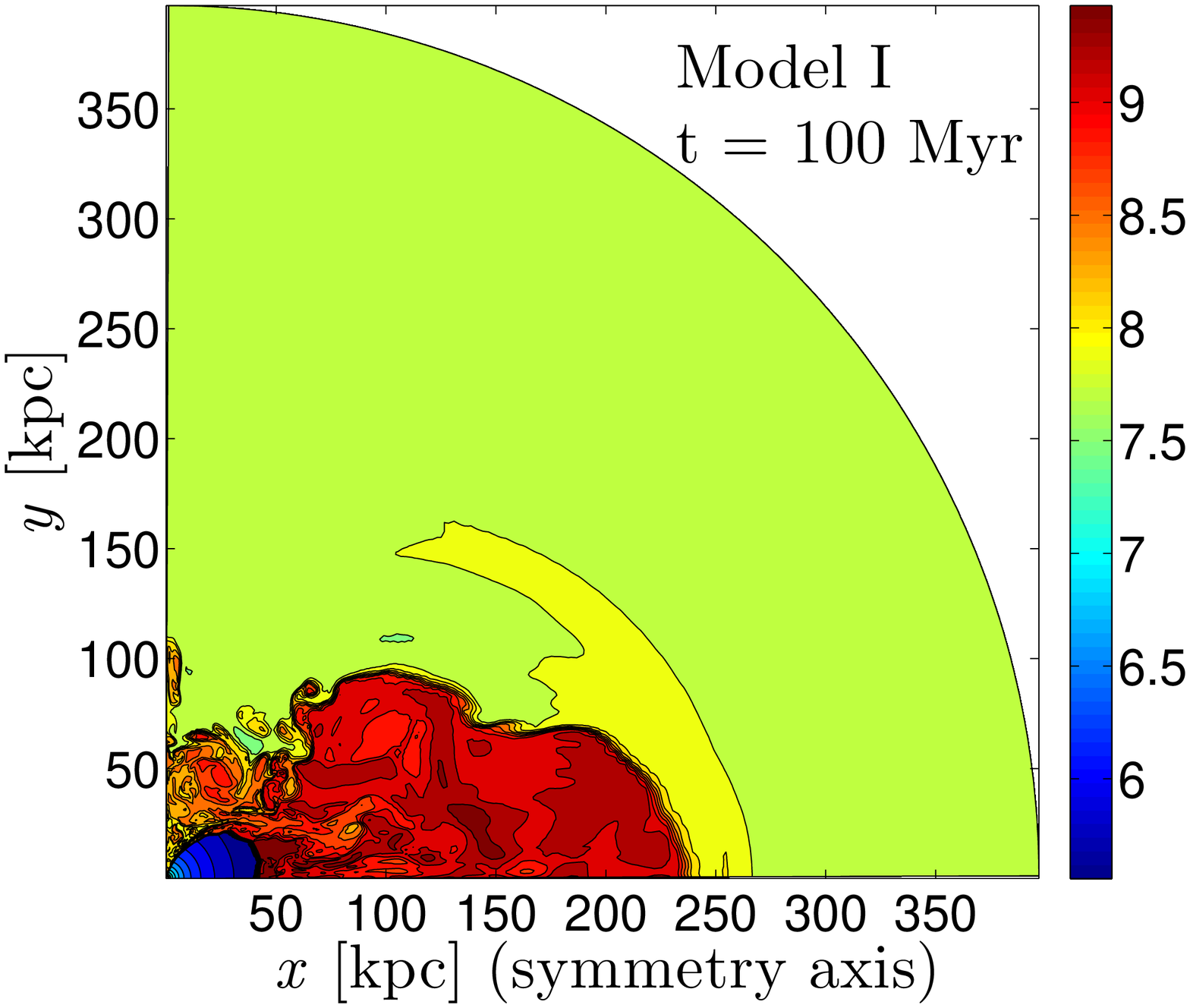}} \\
{\includegraphics[scale=0.5]{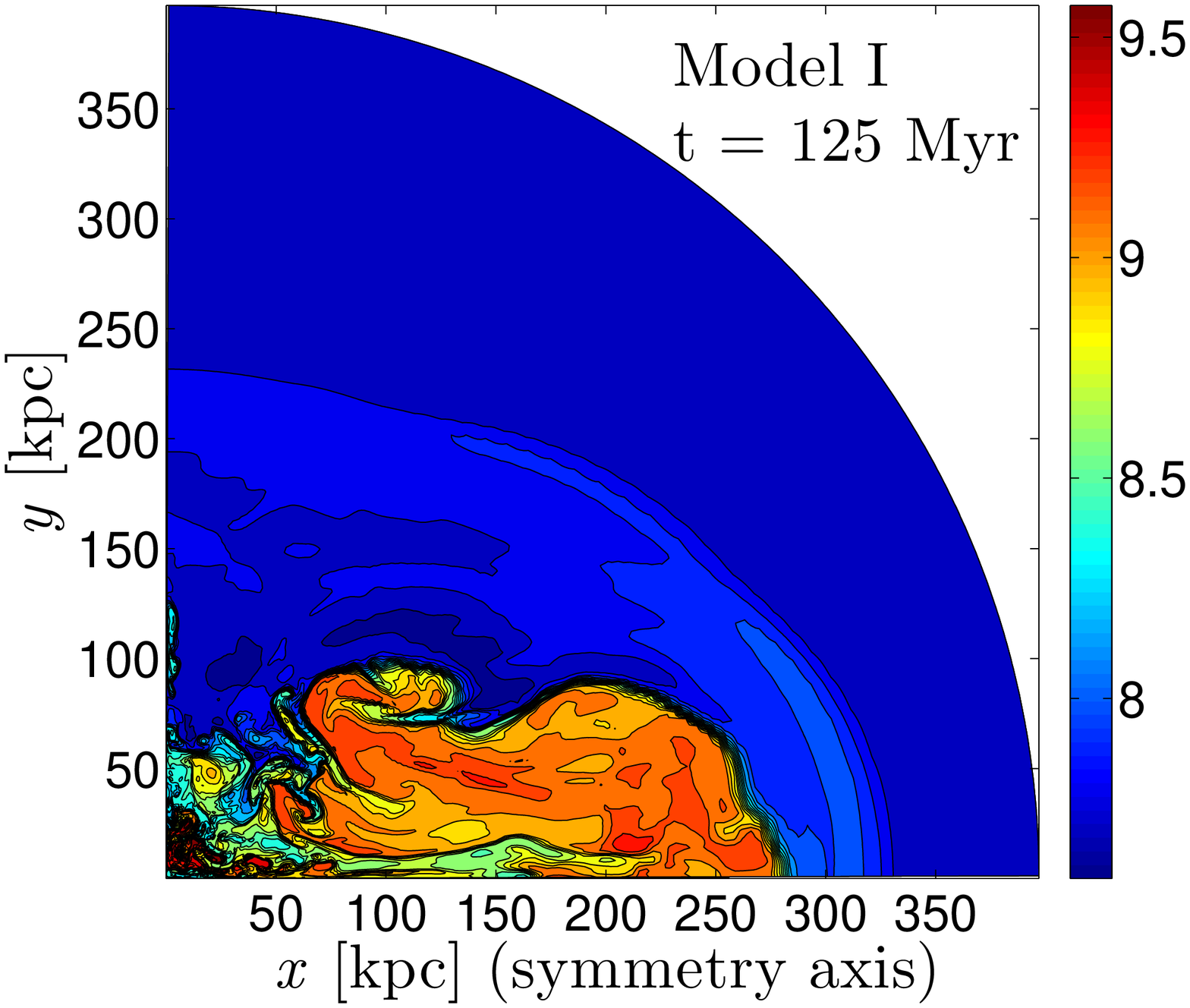}}
\end{tabular}
\caption{Log-temperature map of the bubble inflated by the model I jet at
$t=100\Myr$ and $t=125\Myr$. The temperature is given $\log({\rm K})$.}
\label{fig:M15_temp}
\end{figure}

Figures \ref{fig:M30+M45} and \ref{fig:M30+M45_temp} show density and
temperature maps, respectively (log scales) of the evolution of both a
model II jet (left panels) and a model III (right panels)
jet. The model III jet has a higher velocity and lower mass loss rate compared
to model II, but the same power and the same initial opening angle
(see Table 1 for properties of both jets).
As in model I, in each case the jet was active for $t=100\Myr$.
In both models II and III the jets have inflated low density bubbles attached
to the center. These bubbles are comparable to the one inflated by the model I jet,
though, they are more elongated and the back flow of low density matter
toward the equatorial plane is more pronounced.
The evolution of the bubbles were followed for another $15 \Myr$.
\begin{figure}
\hskip -1.0 cm
\begin{tabular}{cc}
{\includegraphics[scale=0.4]{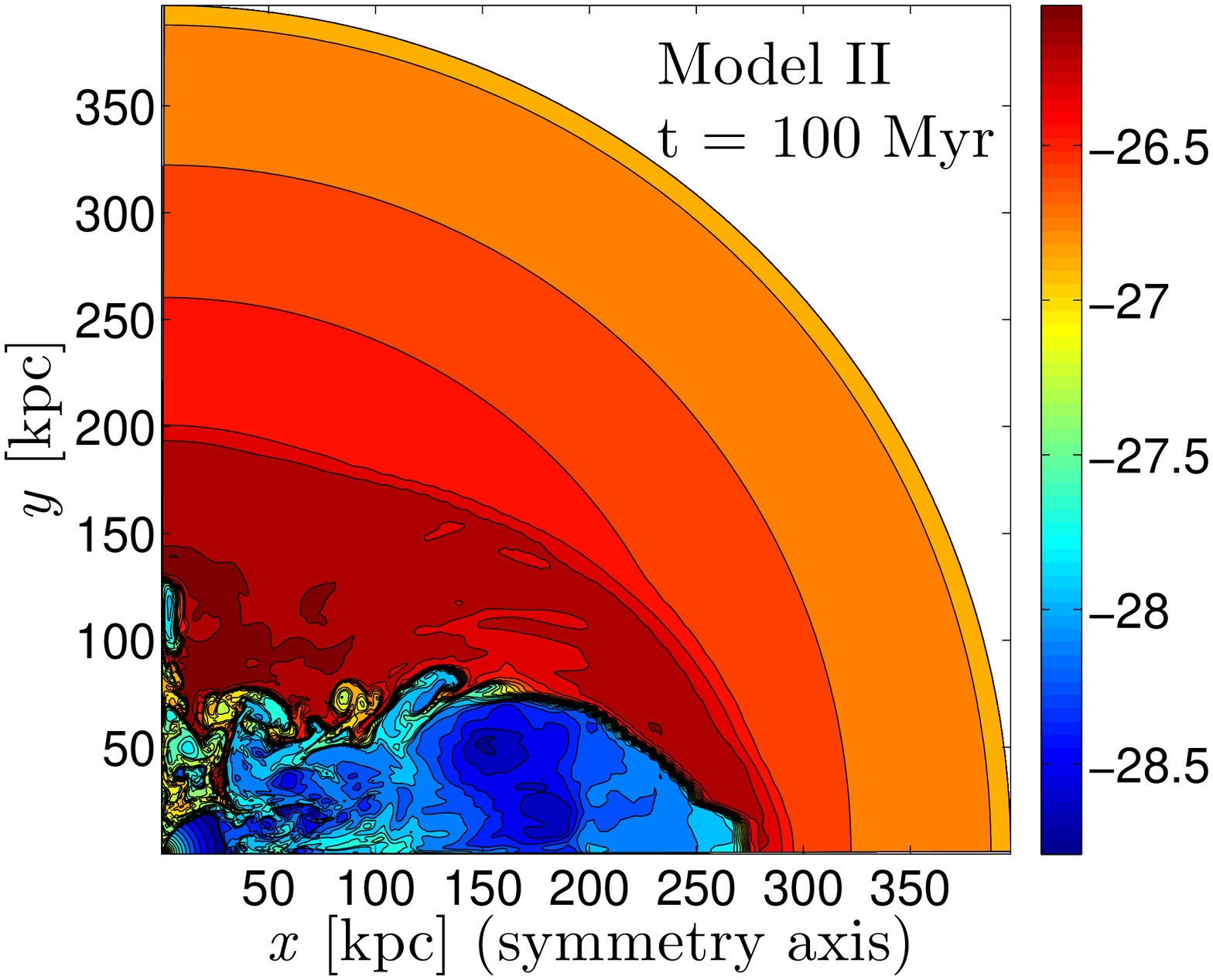}} &
\hskip -0.5 cm
{\includegraphics[scale=0.4]{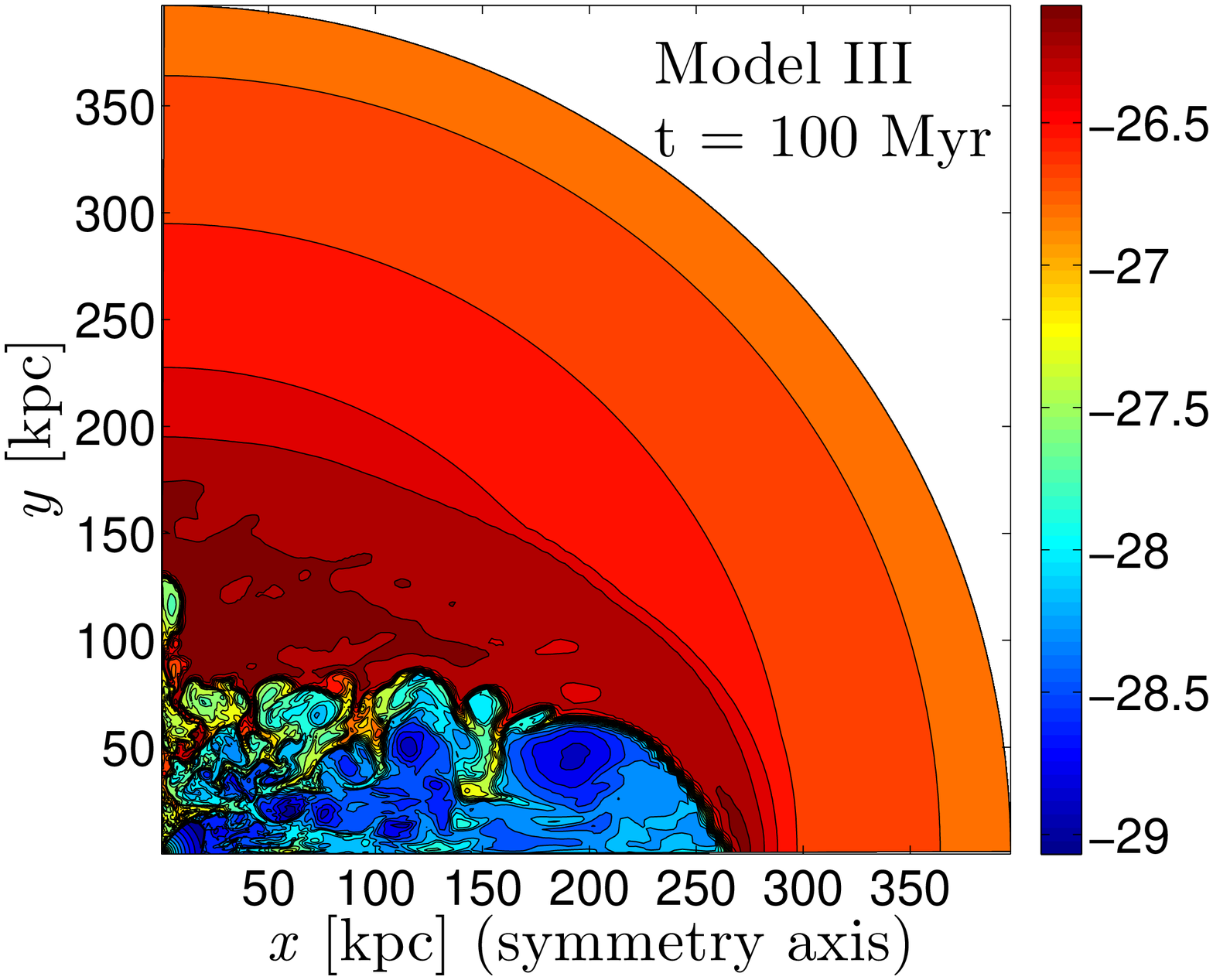}} \\
{\includegraphics[scale=0.4]{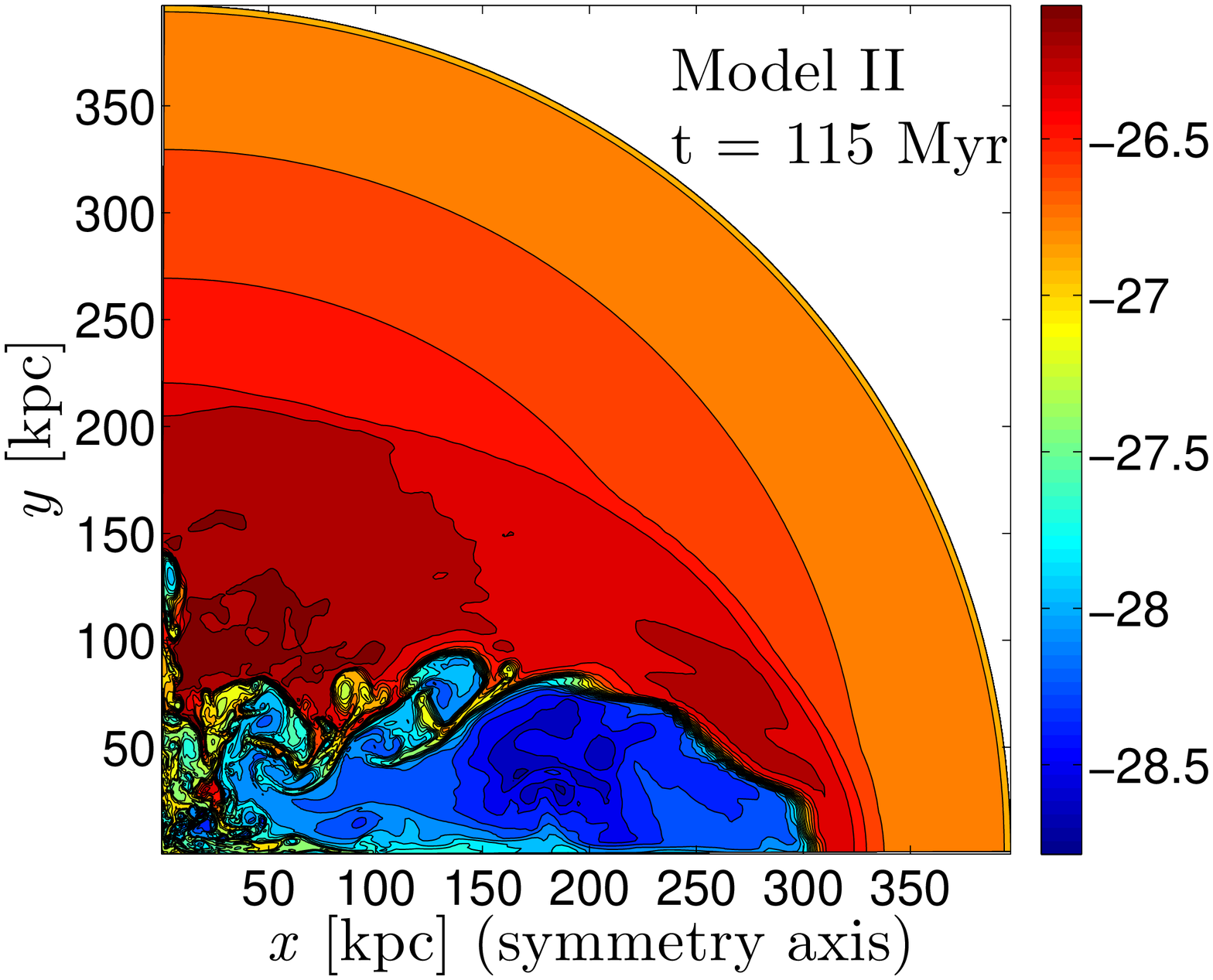}} &
\hskip -0.5 cm
{\includegraphics[scale=0.4]{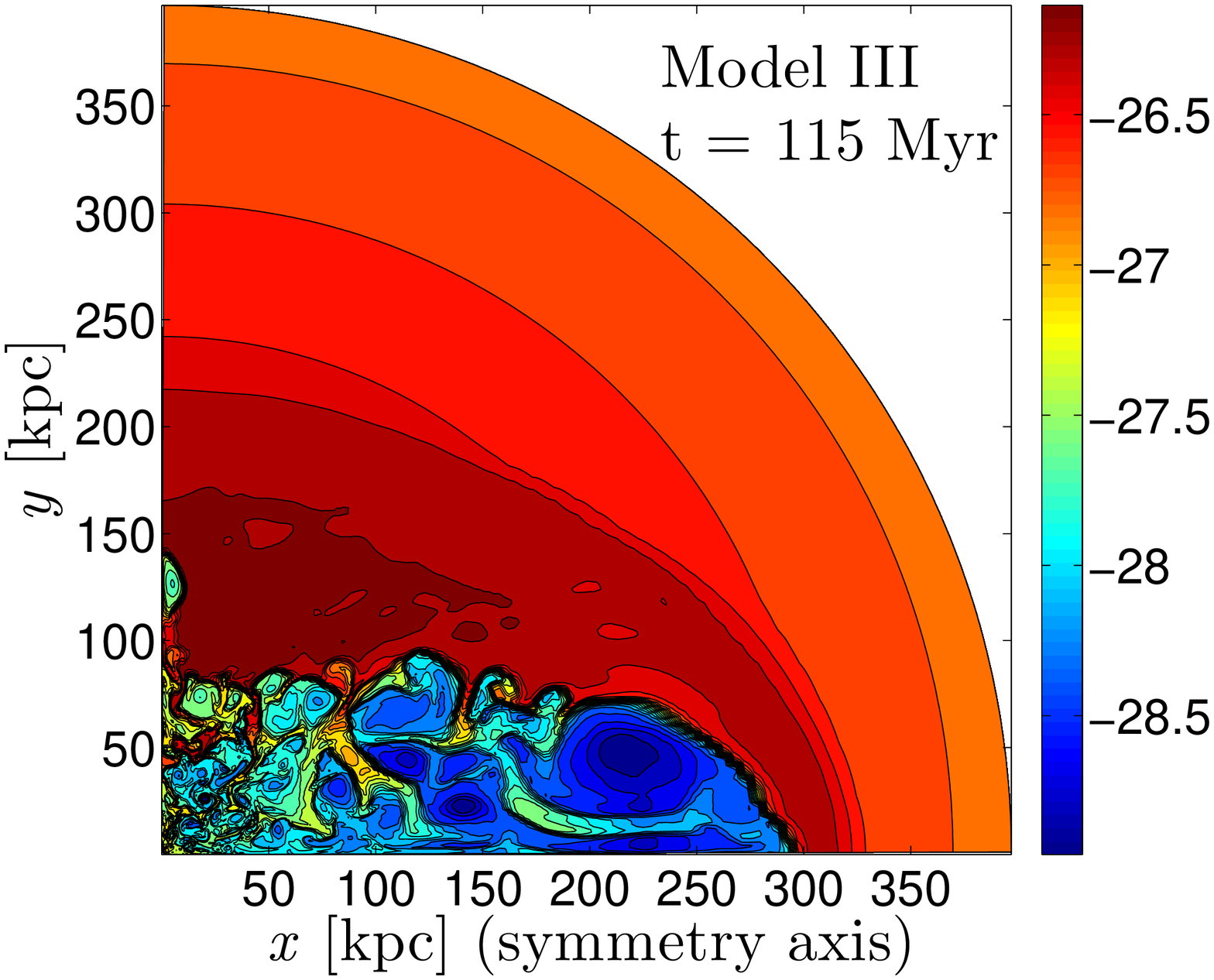}} \\
\end{tabular}
\caption{Log-density maps of the evolution of a model II jet (left panels) and
a model III jet (right panels) shown at two different times (as indicated in
the figures). In both cases the jet was turned off at $t=100\Myr$. Note 
the mixing of dense material into the bubble. The density is given in 
$\log(\gcm)$.}
\label{fig:M30+M45}
\end{figure}
\begin{figure}
\hskip 1.0 cm  
\begin{tabular}{c}  
{\includegraphics[scale=0.5]{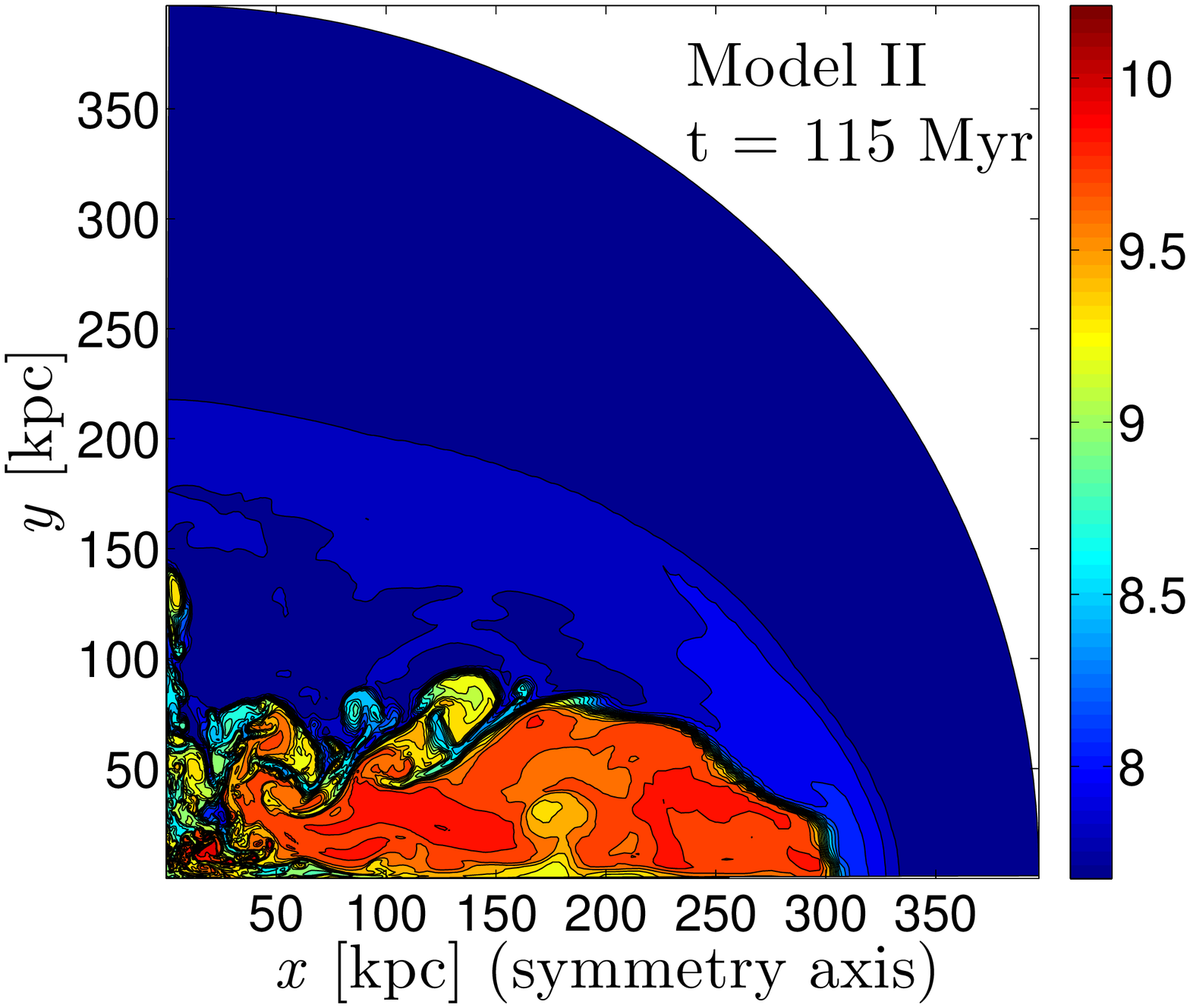}} \\ 
\hskip -0.5 cm
{\includegraphics[scale=0.5]{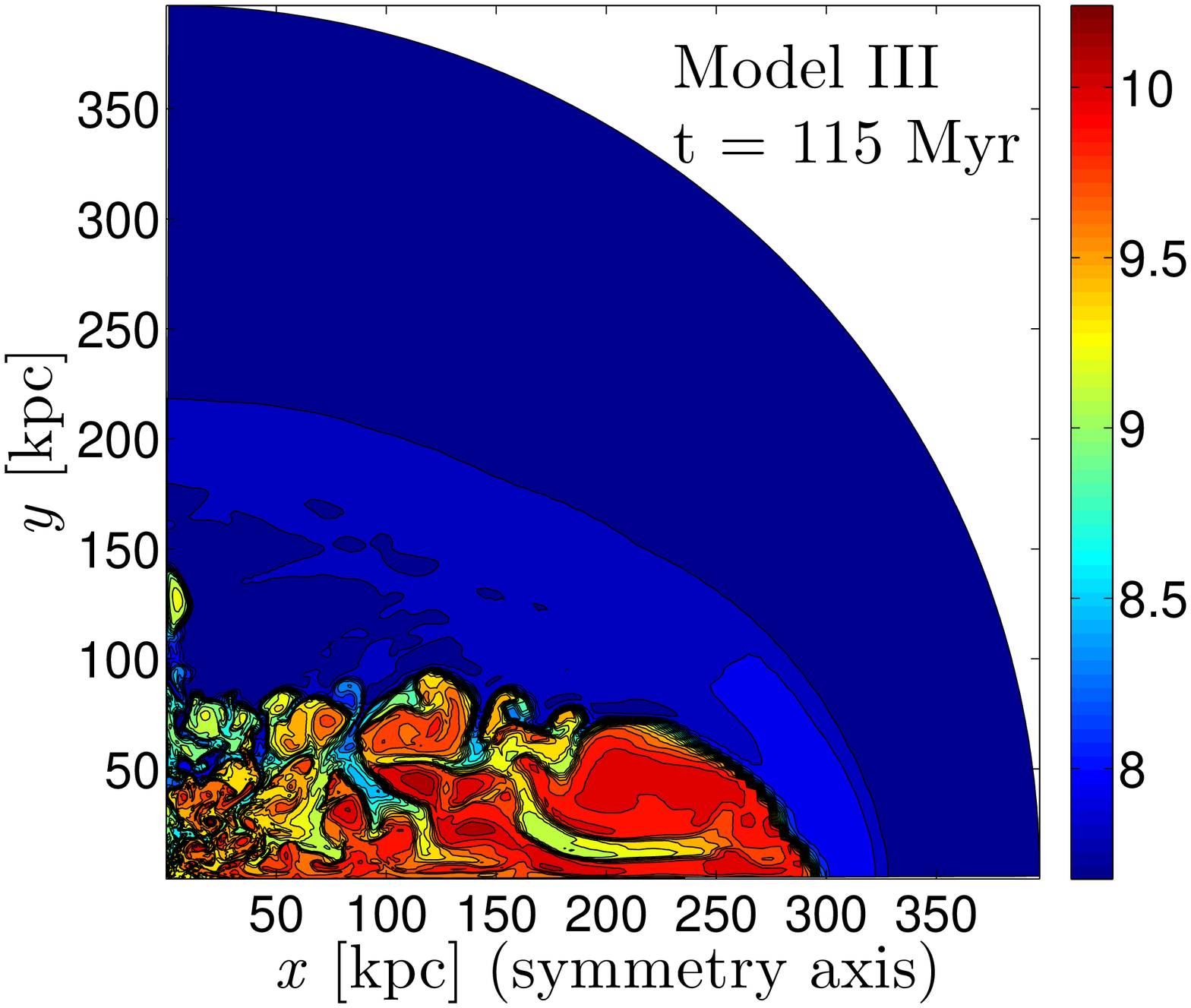}}
\end{tabular}
\caption{Log-temperature map of the bubbles inflated by the model II jet and
a model III jet at $t=115\Myr$. Note the mixing of cold and hot material 
inside the bubble. The temperature is given $\log({\rm K})$.}
\label{fig:M30+M45_temp}
\end{figure}

At $t=115\Myr$, as seen in the bottom left panel of Fig. \ref{fig:M30+M45}, the
model II bubble's center is at a distance of $\sim175\kpc$
from the center. The bubble is elongated with a
major axis of $\sim 250 \kpc$ and a minor axis of $\sim 170\kpc$. The
shock front is elliptical and is located at $\sim330\kpc$ near the symmetry
axis and at $\sim220\kpc$ near the equatorial plane.

At $t=115\Myr$, as seen in the bottom right panel of Fig. \ref{fig:M30+M45},
the model III bubble's center is at a distance of $\sim180\kpc$ from the
cluster center.
It is more elongated than the bubbles inflated in the models I and II.
It's major axis is $\sim 240\kpc$ and it's minor axis is $\sim 150\kpc$.
There is a non-negligible amount of low density matter near the equatorial
plane. The front shock is elliptical, as it is in the other models, and it is located
at $r = 320\kpc$ along the symmetry axis, and $r= 220 \kpc$ in the equatorial plane.

{{{
The bubbles entrain cold gas into them. Most notable is Model III at $t=115 \Myr$,
where tongues of denser gas are seen to penetrate the bubble. It is most likely
that small scale mixing and heat conduction will mix this gas with the hot bubble
gas. Our numerical code cannot handle these small scale processes.
The entrained gas will reduce somewhat the bubble average temperature, but not to
a degree that can be explore with present observations.
In any case, this process is desirable, as it is one of the channels by which
the bubbles can heat the ICM.
}}}

\section{DISCUSSION AND SUMMARY}
\label{sec:summary}

The center of the north and south bubbles in MS0735 reside at distances of
$\sim170\kpc$ and $\sim180\kpc$ from the cluster center, respectively, and
have a radius of $\sim105\kpc$ (Gitti et al. 2007; see Fig. \ref{fig:image} 
here). {{{The actual numbers may vary due to projection effects.}}} 
The model I jet inflates a bubble, that if left to rise buoyantly for 
$25\Myr$, has both size and location
that is in good agreement with the bubbles of MS0735. This is clearly seen in
Figure \ref{fig:M15}. None the less, there are discrepancies between our
results and the observations. A minor discrepancy is in the location of
the shock front. Observations show the shock front to be elliptical with a
semimajor axis of $\sim360\kpc$ and a semiminor axis of $\sim240\kpc$ (Gitti
et al. 2007). While our model I does indeed have an elliptical shock front
with a semiminor axis of $\sim240\kpc$ it's semimajor axis is $\sim330\kpc$.
This discrepancy can be attributed to our cluster density profile that
is not exactly as the real cluster density profile 
{{{or to projection effects (e.g., jet axis not exactly in the plane of the 
sky).}}}

A more serious discrepancy is in the non-negligible low density matter flow
toward the equatorial plane that is present in our simulations but is not
seen in the observations.
Three effects that are not included in our simulations might reduce the presence of the
low density gas near the equatorial plane.
($i$) Including radiative cooling might bring more cooler gas to fall near the equator
replacing some of the hot gas.
($ii$) The turbulent flow near the equator close to the center will stretch magnetic field
lines. In response, the magnetic tension might limit the back flow. It has been shown in
numerical simulations that magnetic fields can prevent small scale mixing and instabilities,
which can prevent some of the backflow.
(e.g., Br\"uggen \& Kaiser 2001; Jones \& De Young 2005; Ruszkowski et al. 2007;
Robinson et al. 2004).
 We note that magnetic fields are not required to stabilized the bubble against
perturbations on large scales (Sternberg \& Soker 2008b); magnetic fields might help
in stabilizing against short wavelength perturbations by supplying tension.
($iii$) Accurate treatment of turbulence and mixing, such as in the subgrid model of
Scannapieco \& Br\"uggen (2008; also Br\"uggen et al. 2009), might prevent some of the
back flow.
These effects will have to be studied in future simulations.   \par

The results of model II and III are less compatible with the observations.
If some of the effects mentioned above are included, it is possible that
model II with a total outflow rate of $\dot{M}_{2j} \simeq 50 M_\odot \yr^{-1}$ can
match observations.
Precessing jets can also allow somewhat faster jets (but still highly subrelativistic).
The bubbles inflated using the models with the faster jets are rather elongated and the amount
of low density matter flowing to the equatorial plane is more substantial.
This supports our preference for rather slow, yet still highly
supersonic, massive jets over the faster and less massive ones, as was also
shown in Sternberg et al. (2007). None the less, the bubbles of MS0735 might
not lie exactly perpendicular to our line of sight. If this is so, then the
bubbles might be more elongated, making the results of models II and III
more agreeable with the observations. It is also plausible that
the bubbles of MS0735 were inflated by precessing jets and not by wide fixed
jets.
{{{ In that case, a narrow jet would rapidly precesses at different angle
around the symmetry axis. The angle of the jet's axis relative to the symmetry
axis of the bubbles would change from zero to $\sim 70^\circ$. }}}
As was shown in Sternberg \& Soker (2008a), rapidly precessing slow
and massive jets inflate bubbles similar to those inflated by SMW jets.
Therefore, even if the jets are rapidly precessing, our model will still
prefer slow and massive jets over faster and less massive ones.  \par

\acknowledgements
We thank John Blondin for his immense help with the numerical code. We thank
{{{ Craig Sarazin, }}} Brian McNamara, Marcus Br\"uggen, Fabio Pizzolato,
{{{ and an anonymous referee }}} for helpful comments.
This research was supported by the Asher Fund for Space
Research at the Technion, and the Israel Science Foundation.


\begin{references}


\reference{} Binney, J. 2004, in The Riddle of Cooling Flows in
     Galaxies and Clusters of Galaxies, Eds. T. Reiprich,
     J. Kempner, and N. Soker, published electronically
     at http://www.astro.virginia.edu/coolflow/proc.php (astroph/0310222)

\reference{} Blanton, E. L., Randall, S. W., Douglass, E. M., Sarazin, C. L., Clarke, T. E.
   \& McNamara, B. R. 2009, ApJ, Letters, in press (arXiv:0904.1610)

\reference{} Blanton, E. L., Sarazin, C. L., McNamara, B. R., \& Wise M. W.
  2001, ApJ, 558, L15

\reference{}  Blondin J.M., Kallman T.R., Fryxell B.A., Taam R.E. 1990, ApJ, 356, 591

\reference{}  Bower, R. G., McCarthy, I. G., \& Benson, A. J. 2008, MNRAS, 390, 1399

\reference{} Br\"uggen, M. \& Kaiser, C. R., 2001, MNRAS, 325, 676

\reference{} Br\"uggen, M., Scannapieco, E., \& Heinz, S. 2009, MNRAS (arXiv:0902.4242)

\reference{} Dupke, R., \& White, R. E. III 2001, High Energy Universe at Sharp Focus: Chandra Science,
ed. E. M. Schlegel \& S. Vrtilek (San Francisco: ASP Conference Series), 51

\reference{} Fabian, A. C., et al.\ 2000, MNRAS, 318, L65 

\reference{} Fioc, M., \&  Rocca-Volmerange, B. 1997, A\&A, 326, 950

\reference{} Gitti, M., McNamara, B. R., Nulsen, P. E. J., \& Wise, M. W.
     2007, ApJ, 660, 1118

\reference{} Gizani, N. A. B., \& Leahy, J. P. 2003, MNRAS, 342, 399

\reference{} Heinz, S., Choi, Y.-Y., Reynolds, C. S., \& Begelman, M. C. 2002, ApJ, 569, L79

\reference{} Jones, T. W. \& De Young, D. S. 2005, ApJ 624, 586

\reference{} Li, L.-X. 2000, PhRvD 61, 4016

\reference{} Livio, M., Ogilvie, G. I., \& Pringle, J. E. 1999, ApJ, 512, 100

\reference{} McNamara, B. R., Kazemzadeh,F., Rafferty, D. A., Birzan, L., Nulsen, P. E. J.,
     Kirkpatrick, C. C., \& Wise, M. W. 2009, ApJ (arXiv:0811.3020)

\reference{} McNamara, B. R., Nulsen, P. E. J., Wise, M. W., Rafferty, D. A., Carilli, C.,
  Sarazin, C. L., \& Blanton, E. L.  2005, Natur, 433, 45

\reference{} McNamara, B. R., Wise, M. W., Nulsen, P. E. J., et al. 2000, ApJ, 534, L135

\reference{} McNamara, B. R., Wise, M. W., Nulsen, P. E. J.,  et al. 2001, ApJ, 562, L149

\reference{} Meier, D. L. 1999, ApJ, 522, 753

\reference{} Meier, D. L. 2001, ApJ, 548, L9

\reference{} Navarro, J. F., Frenk, C. S., \& White, S. D. M. 1996, ApJ, 
462, 563

\reference{} Nulsen, P. E. J., Hambrick, D. C., McNamara, B. R., Rafferty, D.,
  Birzan, L., Wise, M. W., \& David, L. P. 2005, ApJ, 625, L9

\reference{} Pizzolato, F., \& Soker, N. 2005a, ApJ, 632, 821

\reference{} Pizzolato, F., \& Soker, N. 2005b, AdSpR, 36, 762

\reference{} Rizza, E., Loken, C., Bliton, M., Roettiger, K., Burns, J. O., \&
         Owen, F. N. 2000, AJ, 119, 21

\reference{} Robinson, K. et al. 2004, ApJ, 601, 621 

\reference{} Ruszkowski, M., Ensslin, T. A., Br\"uggen, M., Heinz, S., \& Pfrommer, C. 2007, MNRAS,
       378, 662

\reference{} Scannapieco, E., \& Br\"uggen, M. 2008, ApJ, 686, 927

\reference{} Schindler, S., Castillo-Morales, A., De Filippis, E., Schwope, A., \& Wambsganss, J. 2001,
   A\&A 376, L27

\reference{} Simionescu, A., Roediger, E., Nulsen, P.E. J., Br\''uggen, M.,
Forman, W. R., B\''ohringer, H., Werner, N., \& Finoguenov, A. 2009
(arXiv:0810.0271)

\reference{} Soker, N.  2004, A\&A, 414, 943

\reference{} Soker, N. 2006, NewA, 12, 38

\reference{} Soker, N. 2008, NewA, 13, 296

\reference{} Soker, N. \& Pizzolato, F. 2005, ApJ, 622, 847

\reference{} Sternberg, A., Pizzolato, F., \& Soker, N. 2007, ApJ, 656, L5

\reference{} Sternberg, A., \& Soker, N. 2008a, MNRAS, 384, 1327 (2008a)

\reference{} Sternberg, A., \& Soker, N. 2008b, MNRAS, 389, L13  (2008b)

\reference{} Stevens, I. R., Blondin, J. M., \& Pollock, A. M. T. 1992, ApJ, 386, 265

\reference{} Wilson, A. S., Young, A. J., \& Shopbell, P. L. 2000, ApJ, 544, L27

\reference{} Wong, K.-W.  Sarazin, C. L., Blanton, E. L. \& Reiprich, T. H. 2008, ApJ, 682, 155


\end{references}
\end{document}